\documentclass{ws-ijmpa}

\begin{document}

\markboth{Owen Long}
{$CP$ violation, an experimental perspective}

%
\catchline{}{}{}{}{}
%

\title{$CP$ violation, \\ an experimental perspective\footnote{
   Invited plenary talk, DPF meeting, August 2004.}}

\author{\footnotesize OWEN LONG\footnote{\tt owen@slac.stanford.edu}}
\address{Department of Physics, University of California, \\
            Riverside, CA, 92521, USA}

\maketitle


\begin{abstract}

  I present a review of current and near-future
  experimental investigations of $CP$ violation.
  In this review, I cover limits on particle
  electric dipole moments (EDMs) and $CP$ violation
  studies in the $K$ and $B$ systems.  The wealth
  of results from the new $B$ factories provide
  impressive constraints on the CKM quark mixing
  matrix elements.  Current and future measurements
  are focusing on processes dominated by loop diagrams,
  which probe physics at high mass scales in low-energy
  experiments.

\end{abstract}


\section{Introduction}

  The experimental investigation of $CP$ violation seeks to answer
  profound questions about nature.
  One that is often mentioned is, "Why is the universe made
  entirely of matter?"
  The answer to this question must include some kind
  of $CP$ violation\cite{Sakharov}.
  That is, nature can not be symmetric under the combined
  operation of charge conjugation $C$ and parity inversion $P$.
  In the Standard Model, $CP$ violation is due to the irreducible
  phase contained in the 3-generation CKM quark-mixing matrix\cite{CKM}.
  However, the baryon asymmetry in the universe is difficult,
  if not impossible, to explain with $CP$ violation from the
  CKM matrix.
  Neutrino experiments have recently shown that neutrinos are not
  massless particles, as is assumed in the Standard Model.
  This opens the possibility of explaining the baryon
  asymmetry in the universe with $CP$ violation arising
  from flavor mixing in the lepton sector (the so-called theory
  of {\it leptogenesis}\cite{leptogenesis}).
  Another possibility is that the $CP$ violation involved in
  generating the baryon asymmetry is due to physics
  beyond the Standard Model, such as supersymmetry\cite{susy-leptogenesis}.

  The study of $CP$ violation addresses the following more
  general question, "What, if anything, lies beyond the Standard Model of
  particle physics?"
  There are several well-motivated reasons for suspecting that
  the Standard Model is not the final theory of particle
  physics~\cite{NP-arguments}.
  In many extensions of the Standard Model, $CP$ violation is not
  naturally suppressed.
  One can search for new physics by testing the predictions
  the Standard Model makes for a wide variety of
  $CP$ violating observables.
  New $CP$-violating phases from non-Standard Model virtual particles
  in loop corrections can provide a window to discovering new physics at high
  mass scales.
  This approach is most promising when the Standard Model process
  is naturally suppressed.
  In many cases, the Standard Model prediction for the $CP$
  violating observable is quite precise,
  in which case a significant discrepancy would be a clear sign
  of new physics.


  \section{ Particle Electric Dipole Moments }

  A non-zero electric dipole moment (EDM) of a particle, such as
  a neutron, electron, or muon, violates both parity ($P$) and
  time-reversal ($T$) symmetry\cite{ramsey-reviews}.
  This is because the EDM must lie along the direction of the
  spin vector of the particle.
  The energy of a particle in an electric field is given by
  $-d \ \vec{s} \cdot \vec{E}$, where $d$ is the EDM.
  Under parity inversion, the sign of $\vec{E}$ changes, while
  $\vec{s}$ remains the same, since angular momentum is an
  axial vector ($\vec{l} = \vec{r} \times \vec{p}$), thus
  the term of the Hamiltonian given above changes sign under
  parity.
  Similarly, time reversal changes the spin direction while
  leaving $\vec{E}$ the same, changing the sign of the EDM
  term in the Hamiltonian.
  The $CPT$ theorem, a fundamental principle of quantum field
  theory, states that nature is invariant under the combined
  operation of $C$, $P$, and $T$.
  This implies that $T$ violation must be compensated by
  a similar amount of $CP$ violation.

  In the Standard Model, particle EDMs are extremely small,
  since the leading contributions are from 3-loop diagrams.
  A non-zero particle EDM has never been observed.
  The current experimental limits are all orders of magnitude
  above the Standard Model estimates.
  In some new physics scenarios, particle EDMs are greatly enhanced.
  For instance, the leading supersymmetric EDM contributions
  enter at the one-loop level.
  I present the current and future experimental sensitivity
  to the muon and neutron EDM in the remainder of this
  section.


  \subsection{ The EDM of the muon }

     The $g-2$ experiment at Brookhaven has recently
  released a new preliminary upper limit\cite{muonEDM-g-2}
  on the muon EDM.
  The primary mission of the g-2 experiment was to make the
  most precise measurement of the anomalous magnetic moment
  of the muon ($a_\mu \equiv (g-2)/2$).
  The precession of the muon spin, due to the anomalous
  magnetic moment, is given by
  \begin{equation}
    \label{eqn:muon-omega-noedm}
    \vec{\omega_p} =  \frac{e}{m_\mu c} a_\mu \vec{B},
  \end{equation}
  where $\vec{B}$ is the strength of the uniform magnetic
  field of the storage ring.
  The spin precesses in the plane
  of the storage ring ({\it i.e.} $\vec{\omega_p}$ is
  parallel to $\vec{B}$).
  The electrons from muon decay are preferentially emitted
  along the direction of the muon spin.
  This allows the average muon spin to be tracked by
  detecting the position of the decay electrons at several
  locations around the storage ring.

     If the muon has a small, non-zero EDM $d_\mu$, the precession
  vector in Equation~\ref{eqn:muon-omega-noedm} becomes
  \begin{equation}
    \label{eqn:muon-omega-withedm}
    \vec{\omega_p} =  \frac{e}{m_\mu c}
      \left[ a_\mu \vec{B}
      + \frac{1}{2} f \left( \vec{\beta} \times \vec{B} \right) \right],
  \end{equation}
  where $f$ is proportional to the muon EDM ($d_\mu = f \frac{e\hbar}{4mc}$).
  The EDM contribution,
  due to the induced electric field in the muon rest frame,
  is in the radial direction.
  The muon spin precession plane is now slightly tilted with
  respect to the plane of the storage ring.
  The experimental technique is to detect this small tilt by
  monitoring the vertical position of the decay electrons.
  The vertical displacement from $d_\mu$ will
  be {\it exactly} $90^\circ$ out of phase with respect to
  the $g-2$ precession and have the frequency $\omega_p$.
  Significant vertical displacements due to detector
  misalignment combined with coherent betatron oscillations of
  the beam must be carefully removed in the data analysis.

     The result of the analysis~\cite{muonEDM-g-2} is
  $d_\mu = (-0.1 \pm 0.7 \pm 1.2)\times 10^{-19}$~$e$-cm, consistent with zero.
  The 95~\% C.L. upper limit is $d_\mu < 2.8\times10^{-19}$~$e$-cm, which is
  about a factor of 4 lower than the previous limit\cite{muonEDM-previous}.
  This is still $16$ orders of magnitude above the
  Standard Model estimate
  $d_\mu \approx 10^{-35}$ to $10^{-38}$~$e$-cm.
  However, there are a few new-physics
  scenarios~\cite{muonEDM-NP} which
  allow $d_\mu$ to be as large as just 1 to 7 orders of
  magnitude below the current limit.
  The next generation
  experiment\cite{muonEDM-next-generation}
  hopes to extend the sensitivity by another 4 orders of
  magnitude by ``freezing'' the $g-2$ precession with
  a strong radial $E$ field, thus removing the largest
  source of systematic error.
  In this case, the only spin precession is due to the EDM,
  which would produce a significant up-down asymmetry in the
  decay electrons for a $d_\mu$ close to the current limit.


  \subsection{ The EDM of the neutron }

  The current best constraint on the EDM of the neutron
  comes from the RAL/Sussex experiment at
  ILL~\cite{neutronEDM-ILLexpt}.
  It uses the Ramsey resonance technique to measure the
  precession frequency of polarized ultra-cold neutrons 
  in a volume with parallel or antiparallel $E$ and $B$
  fields.
  The precession frequency is given by
  \begin{equation}
  \label{eqn:omegaNeutron}
    \omega_p = \frac{1}{\hbar}\left[
       2 \; \vec{\mu_n} \cdot \vec{B}
       + 2 \; \vec{d}_n \cdot \vec{E} \right]
  \end{equation}
  where $\mu_n$ and $d_n$ are the neutron magnetic and
  electric dipole moments respectively.
  It is easy to see that the difference in
  $\omega_p$ measured with $E$ parallel and
  antiparallel to $B$ gives $d_n$ through the relation
  $\Delta \omega = \frac{1}{\hbar}\left[ 4 \; \vec{d_n} \cdot \vec{E}  \right]$.
  It is essential to continuously monitor the strength
  of the static $B$ field in order to avoid a false EDM signal
  due to drift of the $B$ field between the
  $\omega_p$ measurements with $E$ parallel and
  antiparallel to $B$.
  This was achieved by simultaneously and continuously
  measuring the precession frequency of $^{199}Hg$
  within the same volume as the ultra-cold neutrons.
  Using $^{199}Hg$ as a co-magnetometer removed what was
  the largest source of systematic uncertainty in the
  previous round of experiments.

     The experiment\cite{neutronEDM-ILLexpt} measured
  $d_n = \left( -3.4 \pm 3.9 \pm 3.1 \right)\times 10^{-26}$~$e$-cm,
  consistent with zero.
  The 90\% C.L. upper limit is $\left|d_n\right| < 6.3$~$e$-cm,
  which is 5 orders of magnitude above the Standard Model
  estimates\cite{neutronEDM-SM} of $d_n$
  which are in the range $d_n \approx 10^{-33}$ to
  $2 \times 10^{-31}$~$e$-cm.
  Even though this impressive experimental result is
  consistent with zero, it tells us quite a bit about
  possible new physics scenarios.
  For example, one-loop contributions to the neutron EDM
  in SUSY can give values of order
  $|d_n({\rm SUSY})|\approx\left(\frac{100 \ {\rm GeV}}{m}\right)
    \sin \phi_{A,B} \times 10^{-23}$
  where $m$ is the supersymmetric mass scale and $\phi_{A,B}$ are
  model-dependent phases.
  If the SUSY mass scale is of order 100 GeV, the current
  $d_n$ limit implies that $\phi_{A,B}$ must be small
  (of order $10^{-3}$)~\cite{YNir-review}.

     The $d_n$ limit also address a long-standing mystery of
  the Standard Model -- the so-called ``strong CP problem.''
  The most general QCD lagrangian contains a term that would
  give rise to a $d_n$ of order
  $|d_n({\rm QCD})| \approx 3 \times 10^{-16} \; \bar{\theta} $.
  The Standard Model provides no explanation for why
  $\bar{\theta}$ must be so small (at least $10^{-10}$).
  Pecci and Quinn proposed a solution\cite{pecceiQuinn},
  which adds a new symmetry to the Standard Model and
  predicts the existence of a new particle (the axion),
  which has yet to be observed.

     The proposed next-generation LANSCE neutron EDM
  experiment\cite{neutronEDM-next-generation} hopes to
  increase the $d_n$ sensitivity by another 2 orders of
  magnitude.


  \section{ $CP$ Violation in the Standard Model }

     $CP$ violation in the Standard Model is due to the irreducible
  complex phase within the 3-generation, CKM\cite{CKM} quark-mixing
  matrix\footnote{Now that we know that neutrinos are not massless,
  a complex phase within the neutrino mixing matrix provides
  another mechanism for $CP$ violation.  However, only $CP$ violation
  arising from the CKM matrix is relevant for the rest of this
  discussion.}.
  The electroweak coupling strength in the reaction $W^+\rightarrow q_i \bar{q}_j$
  is proportional to the CKM matrix element $V_{ij}$ where
  $q_i = (u, c, {\rm or}\ t)$ and $\bar{q}_j = (\bar{d}, \bar{s}, {\rm or}\ \bar{b})$.
  For the $CP$-conjugate reaction  $W^-\rightarrow \bar{q}_i q_j$, the CKM matrix
  element is replaced by its complex conjugate $V^*_{ij}$.
  If $V_{ij}$ has a non-trivial phase (not 0 or $\pi$), this phase
  violates $CP$.
  The $CP$ violating phase is only observable through the
  quantum-mechanical interference of at least two amplitudes
  which have both non-zero $CP$ conserving and $CP$ violating
  relative phase differences.

\begin{figure}
  \begin{center}
  \psfig{file=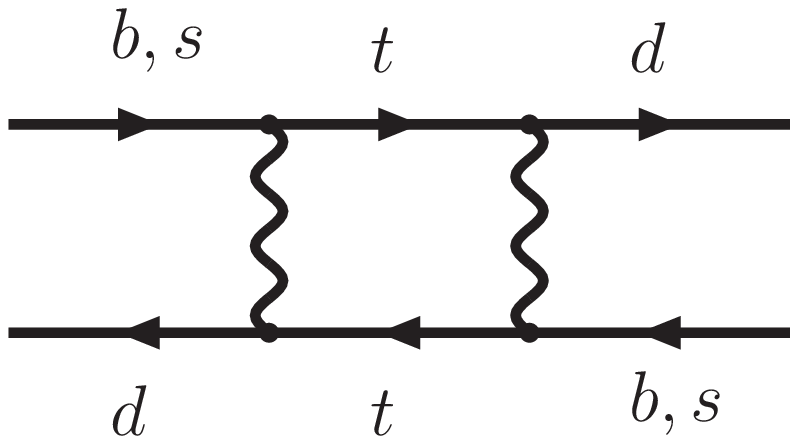,width=4cm}
  \hspace{1cm}
  \psfig{file=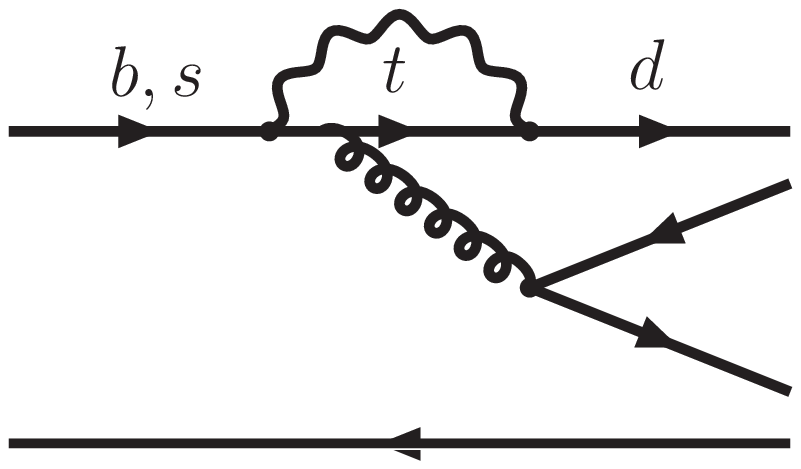,width=4cm}
  \end{center}
  \caption{ The box diagram on the left is an amplitude relevant
  for neutral $B$ and $K$ meson mixing.
  The penguin diagram on the right describes an effective
  flavor-changing neutral current for some rare $K$ and $B$
  decays.  The amplitudes are proportional to the phase
  of $V_{td}^*$ (squared for the box diagram), due to
  the presence of the virtual top quark in the loop.  }
  \label{fig:box-and-peng}
  \end{figure}

     Wolfenstein introduced a very useful parameterization of the
  CKM matrix\cite{Wolfenstein} in terms of 4 fundamental
  parameters ($\lambda, A, \rho, \eta$),
  \begin{equation}
  \label{eqn:ckm}
     V = \left( \begin{array}{ccc}
                  V_{\rm ud}  &   V_{\rm us}  &  V_{\rm ub}  \\
                  V_{\rm cd}  &   V_{\rm cs}  &  V_{\rm cb}  \\
                  V_{\rm td}  &   V_{\rm ts}  &  V_{\rm tb}  \end{array} \right)
 \approx \left(  \begin{array}{ccc}
   1 - \lambda^2/2             &      \lambda        &  A\lambda^3(\rho - i \eta)  \\
   - \lambda                   &   1 - \lambda^2/2   &         A \lambda^2         \\
A\lambda^3(1 - \rho - i \eta)  &    - A \lambda^2    &             1иииииииииииииии
   \end{array} \right) \ \ + \ \ {\cal O}(\lambda^4),
  \end{equation}
  which is an expansion in powers of $\lambda \equiv \sin \theta_c \approx 0.22$,
  where $\theta_c$ is the Cabibbo angle.
  In this parameterization, two important features are evident:  the most
  off-diagonal elements ($V_{ub}$ and $V_{td}$) 1) are the smallest and
  2) contain the non-trivial phase information.
  The phase of $V_{ub}$ can be experimentally probed in $B$ rare
  charmless decays.
  The phase of $V_{td}$ enters through loop diagrams, such as the box
  diagrams describing $B^0$ or $K^0$ mixing or penguin diagrams
  shown in Figure~\ref{fig:box-and-peng}.

  \section{ $CP$ Violation in the Kaon System }

     $CP$ violation was discovered 40 years ago in a famous
  experiment\cite{cronin-fitch} at the Brookhaven National Laboratory where
  the decay $K_L\rightarrow \pi^+\pi^-$ was observed for the
  first time.
  The $\pi^+\pi^-$ final state is $CP$ even and is the dominant decay
  of the $K_S$ meson.
  If $CP$ is conserved, only one of either the $K_S$ or the $K_L$ can
  decay to $\pi^+\pi^-$, but not both\cite{ktopipi-notboth}.

     This kind of $CP$ violation is called indirect $CP$ violation
  or $CP$ violation in mixing.
  It is due to the interference in $K^0 \leftrightarrow \bar{K}^0$ mixing
  between the amplitudes describing transitions through
  virtual intermediate states (e.g. the box diagram) and on-shell
  intermediate states (e.g. $K^0 \rightarrow \pi \pi \rightarrow
  \bar{K}^0$).


  \subsection{ Direct $CP$ violation }

  The second important kind of $CP$ violation in the kaon system does
  not depend on $K^0 \leftrightarrow \bar{K}^0$ mixing.
  If the $CP$ violation is due to at least two decay amplitudes
  directly interfering, it is called ``direct $CP$ violation.''
  Establishing this kind of $CP$ violation has been the focus of
  a series of dedicated experiments.

    The direct-$CP$ violating observable is the double ratio
  \begin{equation}
    \frac{ \Gamma\left( K_L \rightarrow \pi^+ \pi^- \right) /
           \Gamma\left( K_S \rightarrow \pi^+ \pi^- \right) }
         { \Gamma\left( K_L \rightarrow \pi^0 \pi^0 \right) /
           \Gamma\left( K_S \rightarrow \pi^0 \pi^0 \right) } =
     \left| \frac{\eta_{+-}}{\eta_{00}} \right|^2
     \approx 1 + 6\; {\rm Re}\; (\epsilon'/\epsilon)
  \end{equation}
  The NA48\cite{NA48} and KTeV\cite{KTeV} recently published
  their results that removed doubt about whether
  ${\rm Re} \; ( \epsilon' / \epsilon )$ is non-zero.
  The latest world average\cite{PDG} is
  ${\rm Re} \; ( \epsilon' / \epsilon ) = (16.7 \pm 2.3) \times 10^{-4}$.
  This result rules out the superweak theory of $CP$
  violation\cite{superweak}, where $CP$ violation is entirely
  from neutral meson mixing.
  The measured value of ${\rm Re} \; ( \epsilon' / \epsilon )$
  is consistent with the Standard Model calculations, however
  large theoretical uncertainties from the calculation of hadronic
  matrix elements prevent this from being a precision test.


  \subsection{ The Golden Mode -- $K_L\rightarrow \pi^0 \nu \bar{\nu}$ }

  A decay mode that is the focus of much current and future experimental
  effort is the so-called ``golden mode'' $K_L\rightarrow \pi^0 \nu \bar{\nu}$.
  The decay amplitudes are dominated by electroweak loop diagrams.
  The hadronic physics can be calibrated with the common $K^+\rightarrow
  \pi^0 e^+ \nu$ decay.
  This makes the calculation of the branching fraction extremely
  reliable.
  The branching ratio is proportional to ${\rm Im}\; (V_{td} V^*_{ts}) =
  \eta A^2 \lambda^5$, thus giving access to the Wolfenstein parameter
  $\eta$.
  Some new physics scenarios can enhance the branching fraction by up
  to 2 orders of magnitude\cite{Kpi0nunu-susy} above the Standard Model
  prediction\cite{Kpi0nunu-sm} of $(3.0 \pm 0.6)\times 10^{-11}$.

     Observing this decay is extremely challenging.
  The $K_L$ can't be fully reconstructed due to the unobserved neutrinos.
  The main background is from the decay $K_L \rightarrow \pi^0 \pi^0$
  has a branching fraction is $10^7$ times larger than $K_L\rightarrow
  \pi^0 \nu \bar{\nu}$.
  The two additional photons from the extra $\pi^0$ in this mode must
  be detected with extremely high efficiency in order to veto the
  $\pi^0 \pi^0$ background.

     The best direct experimental limit on the branching fraction
  is from KTeV\cite{ktev-pi0nunu}, where they used
  $\pi^0\rightarrow e^+e^-\gamma$.
  They found ${\cal B}(K_L \rightarrow \pi^0 \pi^0) < 5.9 \times 10^{-7}$
  at 90\% C.L., which is well above the predicted value.
  An indirect limit can be derived \cite{grossman-nir-pinunu}
  using the branching fraction from
  the related decay $K^+\rightarrow \pi^+ \nu \bar{\nu}$.
  Using the latest results from the E949 experiment at BNL\cite{e949}
  gives
  \begin{eqnarray*}
    {\cal B}(K_L\rightarrow \pi^0 \nu \bar{\nu}) & < & 4.4 \times
       {\cal B}(K^+\rightarrow \pi^+ \nu \bar{\nu}) \\
    {\cal B}(K_L\rightarrow \pi^0 \nu \bar{\nu}) & < & 1.7 \times 10^{-9}
    \ 90\% \ {\rm C.L.},
  \end{eqnarray*}
  which is only 2 orders of magnitude to the predicted value.
  The E391a experiment at KEK\cite{e391a} is the first dedicated experiment
  for measuring the $K_L\rightarrow \pi^0 \nu \bar{\nu}$ branching
  fraction.
  It's considered a pilot project, since the estimated sensitivity
  is a factor of a few above the Standard Model prediction.
  Using the data that were taken earlier this year, they expect
  to have a sensitivity of about $4 \times 10^{-10}$.
  We eagerly await the E391a results, since some new physics
  models\cite{Kpi0nunu-susy}
  allow for an enhancement of the branching fraction which could
  drive it up to or above the E391a sensitivity.


  \section{ $CP$ Violation in the $B$ System }

  The $B$ system is an excellent place to test the Standard Model
  description of $CP$ violation.
  Unlike the kaon system, the observable $CP$ asymmetries are large (of order 1).
  The theoretical uncertainty in calculation of the expected $CP$
  asymmetry is extremely small in some special cases (about 1\% or less),
  allowing for precision tests of the theory.
  Two new asymmetric-energy $B$ factories were built specifically for
  making these measurements -- the PEP-II storage ring with the Babar
  experiment at SLAC and the KEKB storage ring with the Belle experiment
  at KEK.
  Each experiment has accumulated data samples of well over 200 million
  $B\bar{B}$ events over the last 3.5 years.
  As I will describe below, these enormous datasets have taken our
  understanding of $CP$ violation in the Standard Model to a new
  level of precision.

\begin{figure}
  \begin{center}
  \psfig{file=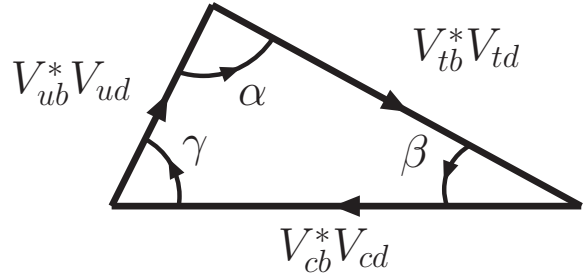,width=8cm}
  \end{center}
  \caption{ The Unitarity Triangle from the CKM unitarity
    constraint
  $V_{ub}^* V_{ud} + V_{cb}^* V_{cd} + V_{tb}^* V_{td} = 0$.
  $CP$ asymmetries in $B$ decays are sensitive to the angles
  of the triangle.
  }
  \label{fig:ut}
  \end{figure}

     A unitarity constraint from the $1^{\rm st}$ and $3^{\rm rd}$ columns
  of the CKM matrix provides a geometrical construction that's useful
  for relating observables in the $B$ system to the fundamental CKM
  parameters.
  This is the so-called ``Unitarity Triangle'', shown in
  Figure~\ref{fig:ut}, which graphically represents the constraint
  $V_{ub}^* V_{ud} + V_{cb}^* V_{cd} + V_{tb}^* V_{td} = 0$.
  The interior angles\footnote{ The Babar collaboration uses the
  $\beta, \alpha, \gamma$ naming convention for the Unitarity Triangle
  angles, while the Belle collaboration uses $\phi_1, \phi_2, \phi_3$.
  I will use the $\beta, \alpha, \gamma$ convention in this note.}
  of the triangles can be measured with
  $CP$ asymmetries.
  In the following sections, I will describe the current status of
  measuring the angles of the triangle.

  \subsection{ Time-dependent $CP$ asymmetries at the asymmetric $B$ factories}

     For final states that can be reached by both $B^0$ and
  $\overline{B}^0$ decay, $B^0\leftrightarrow \overline{B}^0$ mixing
  provides two amplitudes with different CKM phases that can
  interfere with each other -- a necessary requirement for $CP$ violation
  to occur.
  The time-dependent $CP$ asymmetry is defined as
  \begin{equation}
  \label{eqn:acp-def}
    A_{CP}(f;t)  \equiv \frac{N(\overline{B}^0(t) \rightarrow f) -
                              N(          B^0(t)  \rightarrow f) }
                             {N(\overline{B}^0(t) \rightarrow f) +
                              N(          B^0(t)  \rightarrow f) }
  \end{equation}
  where the notation $B^0(t)$ means that the meson was known
  to be (or ``tagged'' as) a $B^0$, as opposed to a $\overline{B}^0$, at $t=0$.
  The $B^0$ and $\overline{B}^0$ from the decay of the $\Upsilon(4S)$ must
  remain flavor-antisymmetric, even while undergoing $B^0\leftrightarrow
  \overline{B}^0$ mixing, due to Bose-Einstein statistics.
  This means that the relevant time in Eqn.~\ref{eqn:acp-def} is the
  time {\em difference} $\Delta t$ between the two $B$ decays, since
  they must be in a flavor-opposite state at $\Delta t=0$.
  Charged decay products of the $B$ that does not decay to $f$ in the
  event are used to infer the flavor of both $B$ mesons at $\Delta t=0$.

     In general, the $A_{CP}(f;\Delta t)$ takes the form
  \begin{equation}
  \label{eqn:acp-lambda}
    A_{CP}(f;\Delta t)  =  \frac{2 \; {\rm Im} \lambda_f}
                                  { 1 + |\lambda_f|^2}
                                \ \sin \Delta m_d \Delta t
                            - \frac{1 - |\lambda_f|^2}
                                   {1 + |\lambda_f|^2}
                                \ \cos \Delta m_d \Delta t \\
  \end{equation}
  \begin{equation}
   S_f \equiv \frac{2 \; {\rm Im} \lambda_f}
                   { 1 + |\lambda_f|^2} \ , \ \ \
   C_f \equiv \frac{1 - |\lambda_f|^2}
                   {1 + |\lambda_f|^2}.
  \end{equation}
  The parameter $\lambda_f$, in the Standard Model, is given by
    $\lambda_f \equiv e^{-i2\beta} \; \bar{A_f} / A_f$, where
  $A_f$ ($\bar{A}_f$) is the amplitude for the $B^0$ ($\overline{B}^0$)
  to decay to $f$.
  This parameter should not be confused with the Wolfenstein parameter $\lambda$.

     The distance between the decay points of the two $B$ mesons
  must be observable in order to reconstruct $\Delta t$ for an event.
  This is the motivation for boosting the $\Upsilon(4S)$ frame in the
  lab by colliding $e^+$ and $e^-$ beams of unequal energies.
  To a good approximation, $\Delta t$ is simply proportional to the longitudinal
  separation between the $B$ decay vertices along the beam boost direction
  ($\Delta t \approx \Delta z/\beta \gamma c$).


  \subsection{ The first precision test of $CP$ violation -- $\sin2\beta$ from $J/\psi K_S$}

     The decay $B^0 \rightarrow J/\psi K_S$ is very special for two reasons.
  First, it is a $CP$ eigenstate with a relatively large branching fraction.
  Second, and more importantly, only a single CKM phase appears in the
  leading decay amplitudes.
  The Standard Model predicts, to within about 1\%\cite{clean-jpsiks},
  $|\overline{A}_{J/\psi K_S}/A_{J/\psi K_S}|=1$,
  $\lambda_{J/\psi K_S} = - e^{-i2\beta}$, and
  \[   S_{J/\psi K_S} = \sin 2\beta \ , \ \ C_{J/\psi K_S} = 0.  \]
  These relations hold for many charmonium $K_S$ final states and for $J/\psi K_L$
  with $S_{J/\psi K_L} = - S_{J/\psi K_S}$.

     Out of all time-dependent $CP$ asymmetry measurements at the new
  $B$ factories, the measurement of $\sin 2 \beta$ from $J/\psi K_S$
  has by far the largest and cleanest sample of signal events.
  The average of the latest results from Babar and
  Belle\cite{sin2beta-babar-and-belle}
  gives $\sin 2\beta = 0.726 \pm 0.037$.
  Indirect constraints on $\beta$ from, $|V_{ub}/V_{cb}|$, $\epsilon_K$,
  $\Delta m_d$, and the limit on $\Delta m_s$ restrict $\beta$ to be
  the range of $[13^\circ,31^\circ]$ at the 95\% C.L.\cite{CKMfitter}.
  One of the 4 solutions for $\beta$ from $\sin 2\beta$ gives
  $\beta = (23.3 \pm 1.6)^\circ$.
  This impressive agreement between the indirect constraints, which
  do not involve $CP$ violation in the $B$ system, and the
  direct measurement led Yosef Nir to conclude\cite{YNir-review}
  ``the Kobayashi-Maskawa mechanism of $CP$ violation has
  successfully passed its first precision test.''


  \subsection{ Looking for New Physics -- the $b\rightarrow s$ penguin modes }

     The $b\rightarrow s$ penguin is the dominant decay amplitude
  for a set of charmless final states, most notably $\phi K_s$.
  The CKM phase in the leading decay amplitudes
  is the same as for $J/\psi K_S$, so to first order these
  modes should also measure $\sin2\beta$
  That is, $S_f \approx -\eta_f \sin 2\beta$ and $C_f \approx 0$, where
  $\eta_f$ is the $CP$ eigenvalue of the final state $f$.
  The comparison of $\sin2\beta$ measured with these $b \rightarrow s$
  penguin modes with the measurement from $J/\psi K_S$ is an intriguing
  test of the Standard Model.
  Some speculate that loop diagrams containing virtual new physics
  (e.g. SUSY) particles may also be contributing to the $b\rightarrow s$
  modes.
  This would, in general, introduce new $CP$ violating phases which could
  significantly alter the $S_f$ and $C_f$ coefficients.

     The first measurements of $S_{\phi K_S}$ were presented at the
  ICHEP 2002 conference in Amsterdam\cite{ICHEP02}.
  Both the Babar and Belle measurements were negative, opposite
  the expected sign, and averaged to $S_{\phi K_S} = -0.39 \pm 0.41$,
  which was $2.7 \: \sigma$ from $S_{J/\psi K_S}$.
  This result generated quite a bit of excitement including dozens
  of phenomenology papers, evaluating $\delta S \equiv S_{\phi K_S} -
  S_{J/\psi K_S}$ in various new physics scenarios.
  At ICHEP 2004 in Bejing\cite{ICHEP04}, updated results were shown from datasets
  more than 3 times larger than those used in the initial measurements.
  The current average is now $S_{\phi K_S} = +0.34 \pm 0.21$, only
  $1.8\:  \sigma$ away from $S_{J/\psi K_S}$.
  However, if one naively averages $-\eta_f S_f$ from all of the $b\rightarrow s$
  penguin measurements\cite{ICHEP04}, one gets $\langle-\eta_f S_f\rangle = +0.42 \pm 0.08$,
  which is $3.6 \: \sigma$ from $S_{J/\psi K_S}$.
  Before concluding that this is evidence for new physics,
  deviations from $-\eta_f S_f = \sin 2\beta$ from sub-dominant
  Standard Model contributions must be carefully
  evaluated\cite{btos-SMpollution}.


  \subsection{ On to the next angle -- $\alpha$ from charmless decays }

  The general technique for measuring the angle $\alpha$ of the Unitarity
  Triangle is to measure the time-dependent asymmetry coefficients
  $S_f$ and $C_f$ of a $CP$ eigenstate $f$ for which the leading decay
  amplitude involves a $b\rightarrow u$ transition.
  In the ideal case, where the CKM phase of the decay amplitude is
  purely that of the $b\rightarrow u$ transition,
  $\lambda_f = \eta_f  e^{-i2\beta} e^{-i2\gamma} = \eta_f  e^{i2\alpha}$,
  $S_f = \eta_f  \sin 2\alpha$, and $C_f = 0$.
  If the tree diagram for $B^0\rightarrow \pi^+ \pi^-$ were the only
  decay amplitude, the $\pi^+\pi^-$ would be ideal for measuring
  $\alpha$.
  Unfortunately, we know that penguin loop diagrams give non-negligible
  contributions to the total $\pi^+\pi^-$ decay amplitude.

     When a second decay amplitude with a different $CP$-violating
  weak phase, such as a penguin contribution, is included, the $\lambda_f$
  parameter is no longer just a pure phase.
  The $\lambda_f$ becomes
  \begin{equation}
     \lambda_f = e^{-i2\alpha} \;
         \frac{ T_f + P_f e^{+i\gamma} e^{i\delta_f} }
              { T_f + P_f e^{-i\gamma} e^{i\delta_f} }
  \end{equation}
  where I have used the unitarity of the CKM matrix to define
  the $T$ and $P$ decay amplitudes in terms of just two
  weak phases.
  The magnitudes of the decay amplitudes $T_f$ and $P_f$ are real
  numbers and $\delta_f$ is relative strong ($CP$-conserving)
  phase between the $T$ and $P$ decay amplitudes.
  The problem now is that $P_f/T_f$ and $\delta_f$ can not be
  reliably calculated in a model-independent way, so they must
  be treated as unknowns that must be determined from experimental
  data.
  This is a significant complication.
  The time-dependent $CP$ asymmetry coefficient $C_f$ can be
  non-zero, due to direct $CP$ violation, and is proportional
  to $\sin \delta_f$.
  The $S_f$ coefficient is $\sqrt{1-C^2_f} \; \sin 2\alpha_{\rm eff}$,
  where $\alpha_{\rm eff}$ goes to $\alpha$ as
  $|P_f/T_f|\rightarrow 0$.

  \subsubsection{ The $\pi\pi$ system }

     Gronau and London pointed out that
  the penguin contribution can be calculated using
  isospin relations from the measured
  decay rates of all of the $\pi\pi$ states ($\pi^+\pi^-$,
  $\pi^0\pi^0$, and $\pi^\pm\pi^0$) for $B^0$ and $\overline{B}^0$
  separately\cite{Gronau-London-Isospin}.
  This is quite challenging experimentally and is limited
  by an 8-fold discrete ambiguity
  in the extraction of $\alpha$ in the range 0 to $\pi$.
  Grossman and Quinn noted a useful inequality derived
  from a geometrical analysis of the $B$ and $\overline{B}$ isospin
  triangles\cite{Grossman-Quinn}
  \begin{equation}
  \label{eqn:grossman-quinn}
    \sin^2(\alpha-\alpha_{\rm eff}) \le
      \frac{ {\cal B}(B^0\rightarrow \pi^0\pi^0) +
             {\cal B}(\overline{B}^0\rightarrow \pi^0\pi^0) }
           { {\cal B}(B^+\rightarrow \pi^+\pi^0) +
             {\cal B}(B^-\rightarrow \pi^-\pi^0) },
  \end{equation}
  which says that if the $\pi^0\pi^0$ branching fraction is small,
  the penguin contribution is small and
  the measured effective value $\alpha_{\rm eff}$ is close to the true
  value of $\alpha$.
  Both the Babar and Belle experiments have seen evidence of the
  $\pi^0\pi^0$ mode\cite{Babar-Belle-2pi0}.
  The current Grossman-Quinn bound gives
  $(\alpha-\alpha_{\rm eff})_{\pi^+\pi^-} < 35^\circ$
  at 90\% C.L.
  which, unfortunately, isn't very restrictive.
  This shortcut can't be used for estimating $\alpha$ from
  $\alpha_{\rm eff}$ -- the full isospin analysis is required
  for the $\pi\pi$ modes.

  \subsubsection{ The $\rho\rho$ system }

     The $\rho^+\rho^-$ mode is not necessarily a $CP$-eigenstate,
  since a vector-vector final state can have $CP$-even ($L=0,2$)
  and $CP$-odd ($L=1$) angular momentum configurations.
  The good news is that the angular analysis of the $\rho^+\rho^-$ mode
  is consistent with full longitudinal
  polarization\cite{babar-rhorho-sandc}, thus it's effectively a
  $CP$-even final state, just like $\pi^+\pi^-$.
  The even better news is that the $\rho^0\rho^0$ mode has not been
  seen\cite{rho0rho0-limits}, implying that the penguin contribution
  is small, unlike the $\pi\pi$ system.  The Grossman-Quinn bound gives
  $(\alpha-\alpha_{\rm eff})_{\rho^+\rho^-} < 11^\circ$ at 90\% C.L.
  implying that $S_{\rho^+\rho^-} \approx \sin 2\alpha$.
  The Babar collaboration has performed the time-dependent $CP$
  analysis of $\rho^+\rho^-$ and finds\cite{babar-rhorho-sandc}
  \begin{eqnarray*}
    C_L(\rho^+\rho^-) & = -0.23 \pm 0.24\; {\rm (stat)} \pm 0.14 \; {\rm (syst)} \\
    S_L(\rho^+\rho^-) & = -0.19 \pm 0.33\; {\rm (stat)} \pm 0.11 \; {\rm (syst)}
  \end{eqnarray*}
  for the longitudinally polarized component.
  The solution, of the two available in the range 0 to $\pi$, closest to
  the value consistent with other CKM constraints gives
  $\alpha = \left[ 96 \pm 10 \; {\rm (stat)} \pm 4 \; {\rm (syst)} \pm 11
  \; {\rm (peng)} \right]^\circ$.

  \subsubsection{ The $\rho\pi$ system }

  The $\rho\pi$ system offers a unique way to resolve tree
  and penguin contributions in a charmless decay.
  Snyder and Quinn noted that a time-dependent $CP$ analysis
  of the $\pi^+\pi^-\pi^0$ Dalitz plane could, in principle,
  determine $\alpha$ without discrete ambiguities\cite{Snyder-Quinn}.
  The regions where the different $\rho\pi$ states overlap in the
  Dalitz plane provide the key information needed to disentangle
  the tree and penguin contributions.

     Thus far, only the Babar experiment has attempted the time-dependent
  Dalitz analysis\cite{Babar-rhopi-Dalitz}, however both Babar and
  Belle\cite{Belle-rhopi}
  have analyzed the $\rho^\pm \pi^\mp$ mode as a quasi-two-body final
  state, by only selecting events near the $\rho^\pm$ resonance and
  removing events in the regions of the Dalitz plot where the
  $\rho\pi$ states interfere.
  One interesting outcome of the quasi-two-body analysis of $\rho^\pm
  \pi^\mp$ is the following direct $CP$ asymmetry
  \begin{equation}
    A^{-+}_{\rho\pi} = \frac{ N(\overline{B}^0 \rightarrow \rho^+\pi^-) -
                              N(          B^0  \rightarrow \rho^-\pi^+) }
                            { N(\overline{B}^0 \rightarrow \rho^+\pi^-) +
                              N(          B^0  \rightarrow \rho^-\pi^+) },
  \end{equation}
  which is the direct $CP$ asymmetry for the diagrams where the
  $\pi$ is from the virtual $W$.
  The two experiments measure
  \begin{eqnarray*}
    {\rm Belle} \ \ \ A^{-+}_{\rho\pi} & = & -0.53
                   \pm 0.29 \;{\rm (stat)}
                   \; ^{+0.09}_{-0.04} \;{\rm (syst)} \\
    {\rm Babar} \ \ A^{-+}_{\rho\pi} & = & -0.47 
                   \; ^{+0.14}_{-0.15} \;{\rm (stat)}
                   \pm 0.06 \;{\rm (syst)},
  \end{eqnarray*}
  The Babar value is derived from the full Dalitz
  analysis\cite{Babar-rhopi-Dalitz}.
  A non-zero value of $A^{-+}_{\rho\pi}$ is evidence of direct $CP$
  violation due to the penguin and tree amplitudes interfering.
  The results of the Babar full Dalitz analysis give
  $\alpha = \left[ 113 \; ^{+27}_{-17} \ {\rm (stat)} \pm 6 \ {\rm
  (syst)} \right]^\circ$ with no discrete ambiguities in the range
  0 to $\pi$.

  \subsubsection{ Combining the modes -- $\alpha$ from charmless $B$ decays }

\begin{figure}
  \begin{center}
  \psfig{file=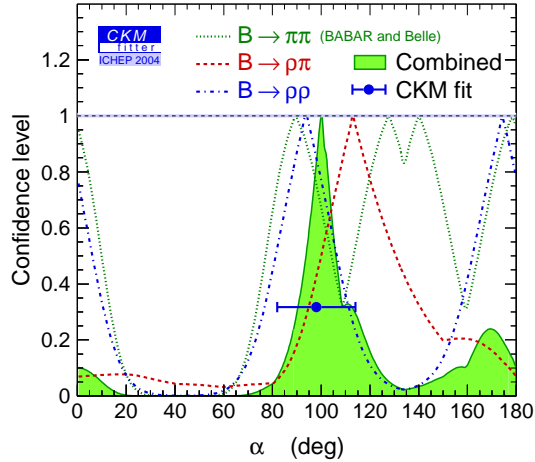,width=7cm}
  \end{center}
  \caption{ Individual constraints on $\alpha$ from
  the analysis of the $\pi\pi$, $\rho\rho$, and $\rho\pi$ systems.
  The solid line is the combined constraint.  The hatched region is
  the constraint from the CKM fit excluding the results shown in
  the figure.
  }
  \label{fig:alpha}
  \end{figure}

  None of the three systems ($\pi\pi$, $\rho\rho$, or $\rho\pi$) gives a
  precise determination of $\alpha$ on it's own, although that may change
  in the future.
  Discrete ambiguities, or false solutions, are a problem, especially
  for $\pi\pi$.
  However, only the true solution will be the same for all three systems
  -- any overlap amongst the false solutions is accidental.
  Figure~\ref{fig:alpha} shows the individual constraints on $\alpha$
  from the three systems and the combined constraint\cite{alpha-fig}.
  Combining all three modes gives $\alpha = \left[ 103 \pm
  11\right]^\circ$, which is in excellent agreement with the allowed
  value from other CKM constraints\cite{CKMfitter2}
  $\alpha_{CKM} = \left[ 98 \pm 16 \right]^\circ$.
  This is another victory for the Standard Model.

  \subsection{ Direct $CP$ violation again -- $A(K^\pm \pi^\mp)$ is non-zero }

  The $K^\pm \pi^\mp$ mode is the main background from $B$ decays
  for $\pi^+\pi^-$, but it's very interesting in its own right.
  The CKM factors for the top quark penguin amplitude are much larger than
  the tree amplitude: $|P/T|_{CKM} = |V_{tb}V_{ts}/V_{ub}V_{us}|
  \approx 1/(0.4 \lambda^2) \approx 50$.
  However, the penguin loop diagram is suppressed with respect to
  the tree, so this somewhat reduces the factor of 50.
  Since this mode is ``self-tagging'' ($K^+\pi^-$ only comes from a
  $B^0$ and $K^-\pi^+$ only comes from a $\overline{B}^0$)
  the only kind of Standard Model $CP$ violation it can exhibit is
  through interfering decay amplitudes, or direct $CP$ violation.
  The current measurements of the asymmetry
  \begin{equation}
    A_{K\pi} = \frac{ N(K^-\pi^+) - N(K^+\pi^-) }
                    { N(K^-\pi^+) + N(K^+\pi^-) }
  \end{equation}
  are
  \begin{eqnarray*}
    {\rm Babar\mbox{\cite{babar-akpi}}} \ \ A_{K\pi} & = & -0.133
                   \; \pm 0.030 \;{\rm (stat)}
                   \; \pm 0.009 \;{\rm (syst)}  \\
    {\rm Belle\mbox{\cite{belle-akpi}}} \ \ \ A_{K\pi} & = & -0.101
                   \; \pm 0.025 \;{\rm (stat)}
                   \; \pm 0.005 \;{\rm (syst)}
  \end{eqnarray*}
  with an average of $A_{K\pi} = -0.114 \pm 0.020$,
  which is 5.7~$\sigma$ from zero.
  Like ${\rm Re} (\epsilon'/\epsilon)$ in the kaon system, this
  establishes the phenomenon of direct $CP$ violation in the $B$
  system.


  \subsection{ Mission impossible? --  $\gamma$ from $B^\pm \rightarrow DK^\pm$ }

  The angle $\gamma$ of the unitarity triangle,
  which is the relative phase of the $b\rightarrow u$ transition
  with respect to the $b\rightarrow c$ transition, is by far the most
  difficult to measure.
  A fairly straightforward and theoretically clean method was
  proposed by Gronau, London, and Weyler\cite{GLW}.
  The decay $B^-\rightarrow D^0 K^-$ proceed via a $b\rightarrow c$
  tree diagram, while the decay $B^-\rightarrow \overline{D}^0 K^-$
  goes through a color-suppressed $b\rightarrow u$ diagram.
  For final states that both the $D^0$ and the $\overline{D}^0$ can
  decay to, these two paths will interfere, thus enabling the
  determination of the relative weak phase $\gamma$.

     A crucial parameter in this method is the relative size of
  the $b\rightarrow u$ and $b\rightarrow c$ amplitudes, which is
  defined $r_B \equiv |A(b\rightarrow u)/A(b\rightarrow c)|$.
  Unfortunately, $r_B$ can't be reliably calculated.
  A rough estimate is $r_B \approx 0.4 \; F_{\rm cs}$, where the 0.4 is
  from the ratio of CKM elements and $F_{\rm cs}$ is an unknown
  color suppression factor, which is expected to be in the range
  of [0.2,0.5].
  This gives an expected range for $r_B$ of [0.1,0.2].

     The original $B\rightarrow DK$ proposal\cite{GLW} was to use $D$ decays to
  $CP$ eigenstates (e.g. $\pi^+\pi^-$) thereby forcing equal $D^0$
  and $\overline{D}^0$ decay amplitudes by construction.
  The problem with this technique is that the interference terms
  are of order $r_B$, which may be small.
  An alternative  $B\rightarrow DK$ method\cite{ADS} uses $D$ decays to flavor-specific
  final states (e.g. $K^-\pi^+$).
  If the dominant $B$ decay ($B^-\rightarrow D^0 K^-$) is combined
  with the suppressed $D$ decay ($D^0\rightarrow K^+\pi^-$) the overall
  amplitudes of the $b\rightarrow u$ and $b\rightarrow c$ paths become
  comparable, thus maximizing the interference effects at the cost
  of a lower overall rate.
  Finally, a hybrid $B\rightarrow DK$ approach was recently proposed\cite{DK-Dalitz}
  where one performs a direct $CP$ violation analysis in the Dalitz
  plane of a 3-body $D$ decay, such as $D\rightarrow K_s \pi^+\pi^-$.
  The resonance structures of the Dalitz plane can be externally
  determined from copious $c\overline{c}$ samples.
  Overlapping resonances from the $D^0$ and $\overline{D}^0$ produce
  ``hot spots'' of relatively large interference with a known strong
  phase variation, which in principle allows for an ambiguity-free
  determination of $\gamma$ in the range 0 to $\pi$.

      Many measurements have been made of $B\rightarrow D^{(*)}K^{(*)}$ decay rates using
  $D$ decays to $CP$ eigenstates\cite{DKmeas-CP}, however the statistical
  errors are still substantial.
  Both the Babar and Belle collaborations have investigated
  $[K^+\pi^-]_D K^-$ and its charge conjugate\cite{DKmeas-ADS}.
  They both see only hints of a signal, although the hint is
  larger for Belle.
  The upper limits on the $[K^\pm\pi^\mp]_D K^\mp$ rate can be translated
  into the following 90\% upper limits on $r_B$: $r_B<0.28$ (Belle),
  $r_B<0.23$ and $r_B^*<0.21$ (Babar) where the last ($r_B^*$) is
  for the $D^*K$ mode.
  These results favor a small value of $r_B$.

     Belle and Babar have also performed the $DK$ Dalitz analysis using
  the decay $D\rightarrow K_s \pi^+\pi^-$ using both $DK$ and $D^*K$
  decays\cite{DKmeas-Dalitz-Belle}$^,$\cite{DKmeas-Dalitz-Babar}.
  Choosing the (single) solution for $\gamma$ in the range 0 to $\pi$,
  the results are
  \begin{eqnarray*}
    {\rm Belle\mbox{\cite{DKmeas-Dalitz-Belle}}} \ \ \ \gamma & = &
       \left[ 77
                   \  ^{+17}_{-19} \;{\rm (stat)}
                   \; \pm 13 \;{\rm (syst)}
                   \; \pm 11 \;{\rm (model)} \right]^\circ \\
    {\rm Babar\mbox{\cite{DKmeas-Dalitz-Babar}}} \ \ \gamma & = &
       \left[ 88
                   \; \pm 41 \;{\rm (stat)}
                   \; \pm 19 \;{\rm (syst)}
                   \; \pm 10 \;{\rm (model)} \right]^\circ,
  \end{eqnarray*}
  where the Belle analysis used the Feldman-Cousins technique,
  while the Babar analysis interpreted their measurement using a
  Bayesian approach.
  The last uncertainty is from the resonance parameters in the
  Dalitz model.
  The large difference in statistical errors is partly due to the
  fact that the Belle (Babar) data favor a large (small) value
  of $r_B$.

      All of the $B\rightarrow DK$ techniques mentioned in this
  section depend on the same set of three unknowns: $r_B$,
  $\gamma$, and $\delta_B$, where $\delta_B$ is the strong phase
  difference between the $b\rightarrow c$ and the $b\rightarrow u$
  decay amplitudes.
  For this reason, it makes sense to combine all of the measurements
  in a global analysis.
  For example, the UTfit group has combined all $B\rightarrow DK$
  data available using a Bayesian analysis\cite{UTfit-gamma}.
  They find
  $r_B = 0.10 \pm 0.04$.
  and
  $\gamma = \left[ 68 \pm 19 \right]^\circ$,
  which is consistent with the
  value from other CKM constraints\cite{UTfit-gamma}
  $\gamma_{CKM} = \left[ 60 \pm 7 \right]^\circ$.

       Other approaches to measuring $\gamma$ include the time-dependent
  analysis of $B^0\rightarrow D^{(*)}\pi$ and $B^0\rightarrow D^{(*)}K^{(*)}$,
  where the asymmetry is proportional to $\sin(2\beta+\gamma)$\cite{sin2bpg}, and
  studying the $B\rightarrow K\pi$ modes\cite{gamma-from-kpi}.


  \subsection{ $B$ system summary }

  The new $B$ factories KEKB and PEP-II and their associated experiments,
  Belle and Babar, have contributed greatly to our understanding
  of $CP$ violation.
  The measurement of $\sin 2 \beta$ from charmonium decays was the
  first precision test of the Standard Model description of $CP$
  violation -- a test it passed with flying colors.
  The $b\rightarrow s$ penguin decay modes show an interesting
  discrepancy with $\sin 2 \beta$ from charmonium decays, which
  may be a hint of new physics.
  Resolving this discrepancy is a high priority for current and future
  experiments.
  Much progress has been made in determining the angle $\alpha$ of the
  unitarity triangle with charmless decays, most notably $\rho^+\rho^-$
  and $\rho\pi$.
  Direct $CP$ violation has been established in the $B$ system with the
  measurement of $A_{K\pi}$.
  Finally, determining the angle $\gamma$ of the unitarity triangle looks
  like it's going to be difficult, as expected -- $r_B$ is not large.


  \section{ Concluding Remarks }

  The Standard Model description of $CP$ violation is remarkably
  successful.
  There are many new and improved experimental constraints and no
  significant discrepancy has been found, although there are some
  interesting anomalies.
  Current and future experiments are now focusing on studying
  processes dominated by loop diagrams in the standard model:
  particle EDMs, $K_L\rightarrow \pi^0\nu\overline{\nu}$, and
  $CP$ asymmetries in penguin-dominated $B$ decays.
  Loop diagrams bring in sensitivity to high virtual mass scales
  where new physics contributions may be significant.
  With experimental advances expected on all fronts, the
  study of $CP$ violation in the next decade promises to be very
  exciting.


\section*{Acknowledgments}

  I would like to acknowledge the friendly cooperation
  of several members of the experimental collaborations
  whose results have been presented here.
  I would especially like to thank Andreas H\"{o}ecker of
  the CKMfitter group\footnote{\tt http://ckmfitter.in2p3.fr}
  and Marcella Bona and Maurizio Pierini of the
  UTFit group\footnote{\tt http://www.utfit.org} for rapidly providing their
  latest CKM analysis results.


\end{document}